\documentclass[twoside,twocolumn,english,prl,showpacs]{revtex4}
\usepackage[T1]{fontenc}
\usepackage[latin1]{inputenc}
\usepackage{graphicx}
\usepackage{amssymb}

\makeatletter

\usepackage{txfonts}

\usepackage{babel}
\makeatother
\begin{document}

\title{Direct extraction of the Eliashberg function for electron-phonon
coupling: A case study of Be(10$\mathbf{\bar{1}}$0)}

\author{Junren~Shi,$^{1}$ S.-J~Tang,$^{2}$ Biao~Wu,$^{1,\,3}$
P.T.~Sprunger,$^{4}$ W.L.~Yang,$^{5,\,6}$ V.~Brouet,$^{5,\,6}$
X.J.~Zhou,$^{5}$ Z.  Hussain,$^{6}$ Z.-X.~Shen,$^{5}$
Zhenyu~Zhang,$^{1,\,2}$ E.W.~Plummer$^{1,\,2}$}

\affiliation{ 
$^{1}$Condensed Mater Sciences Division, Oak Ridge National
Laboratory, Oak Ridge, TN 37831\\
$^{2}$Department of Physics and Astronomy, University of Tennessee,
Knoxville, TN 37996\\
$^{3}$Department of Physics, University of Texas, Austin, TX 78712 \\
$^{4}$Department of Physics and Astronomy, Louisiana State University,
Baton Rouge, LA 70803\\
 $^{5}$Department of Physics, Applied Physics and Stanford Synchrotron
Radiation Laboratory, Stanford University, Stanford, CA 94305\\
$^{6}$Advanced Light Source, Lawrence Berkeley National Laboratory,
Berkeley, CA 94720
}

\begin{abstract}
We propose a systematic procedure to directly extract the Eliashberg
function for electron-phonon coupling from high-resolution
angle-resolved photoemission measurement. The procedure is
successfully applied to the Be(10$\bar{1}$0) surface, providing new
insights to electron-phonon coupling at this surface. The method is
shown to be robust against imperfections in experimental data and
suitable for wider applications.
\end{abstract}


\maketitle 

Electron-phonon coupling (EPC) is the basis for many interesting
phenomena in condensed matter physics such as conventional
superconductivity.  Its possible role in the high-$T_{c}$ cuprates is
also being actively
discussed~\cite{Darmascelli2003,Egami2002,Lanzara2001}. Experimentally,
recent advances in high-resolution (both energy and momentum)
angle-resolved photoemission spectroscopy (ARPES) have stimulated many
studies on EPC in various
systems~\cite{Lanzara2001,Matzdorf1996,Balasubramanian1998,Hengsberger1999,Valla1999,Lashell2000,Rotenberg2000,Tang2002,Luh2002,Ast2002}.
These ARPES measurements usually yield the mass enhancement factor
$\lambda$~\cite{Grimvall}, which characterizes the strength of EPC,
along with some primitive information about its spectral structure
such as the dominant phonon mode. It is highly desirable to take
advantage of these high-resolution data and obtain full
characteristics of EPC.

Theoretically, the full characteristics of EPC are described by the
Eliashberg function
$\alpha^{2}F(\omega;\epsilon,\mathbf{\hat{k}})$~\cite{Grimvall}, the
total transition probability of a quasi-particle from/to the state
$(\epsilon,\mathbf{\hat{k}})$ by coupling to boson (phonon) modes of
frequency $\omega$~\cite{Eiguren2002}. Essentially all physical
quantities related to EPC can be deduced from the function. For
instance, the mass enhancement factor $\lambda$ is related to the
Eliashberg function by~\cite{Grimvall},
\begin{equation}
\lambda=2\int_{0}^{\infty}\frac{d\omega}{\omega}\alpha^{2}
F(\omega;\epsilon_{F},\hat{\mathbf{k}})\,.\label{eq:integral lambda}
\end{equation}

In this Letter, we present a systematic procedure to directly extract
the Eliashberg function from the high-resolution ARPES data. The
Maximum Entropy Method (MEM)~\cite{Gubernatis1991} is employed to
overcome the numerical instability inherent to such
efforts~\cite{Verga2003}.  With Eq.\ref{eq:integral lambda}, our
procedure also gives a reliable estimate of the mass enhancement
factor $\lambda$, without making ad hoc assumptions on phonon model or
requiring low temperature
measurement~\cite{Ast2002,Balasubramanian1998,Hengsberger1999,Lashell2000,Tang2002,Rotenberg2000,Valla1999,Luh2002}.

This procedure is illustrated using new high-resolution ARPES data at
the Be(10$\bar{1}$0) surface. This system is ideal for testing the new
procedure because Be has a broad phonon band ($\sim80\,\mathrm{meV}$),
thereby removing the need for super-high-energy resolution. Also,
results of measurements~\cite{Tang2002,Hofmann1997} and theoretical
calculations~\cite{Lazzeri2000} for this surface have been published,
providing references for comparison.

\begin{figure}
\includegraphics[%
  clip,
  width=1.0\columnwidth]{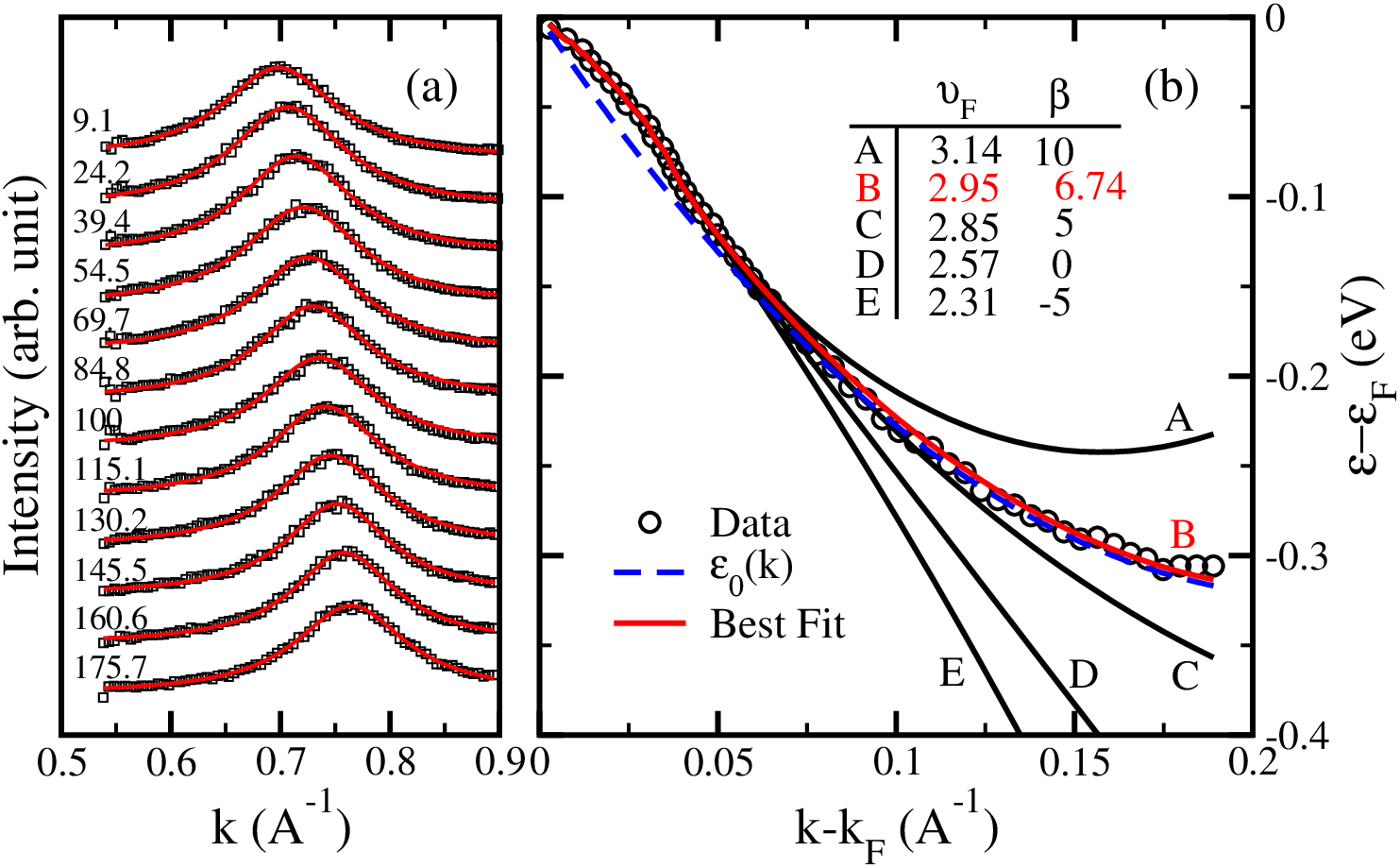}

\caption{\label{cap:photoemmission} (Color online) (a) Momentum
distribution curves (MDCs) of S1 surface state of Be(10$\bar{1}$0)
along $\bar{\mathrm{A}}\bar{\Gamma}$.  The numbers denote the initial
state energies in meV. The solid lines show the Lorentzian
fittings. (b) Quasi-particle dispersion determined from the MDCs
(circles). The solid lines show fittings to the dispersion with the
procedure detailed in the text using different bare quasi-particle
dispersions $\epsilon_{0}(\mathbf{k})$. The parameters $\upsilon_{F}$
($\mathrm{eV}\cdot\mathrm{\AA}/\hbar$) and $\beta$
($\mathrm{eV}\cdot\mathrm{\AA}^{2})$ of $\epsilon_{0}(\mathbf{k})$ are
shown in the inset. The dashed line indicates
$\epsilon_{0}(\mathbf{k})$ that results in the best fit. }
\end{figure}

The photoemission experiments were performed at the Advanced Light
Source (ALS) on Beamline 10.0.1 using a display high-resolution
Scienta 2002 energy analyzer at 24\, eV photon energy with total
energy resolution $10\,\mathrm{meV}$ and angular resolution
$\pm0.15^{\circ}$ in $6\times10^{-11}$ Torr vacuum and at
$T=30\,\mathrm{K}$. The cleaning procedure for the Be(10$\bar{1}$0)
sample was described earlier~\cite{Balasubramanian1998}.

\begin{figure}
\includegraphics[%
  clip,
  width=0.90\columnwidth]{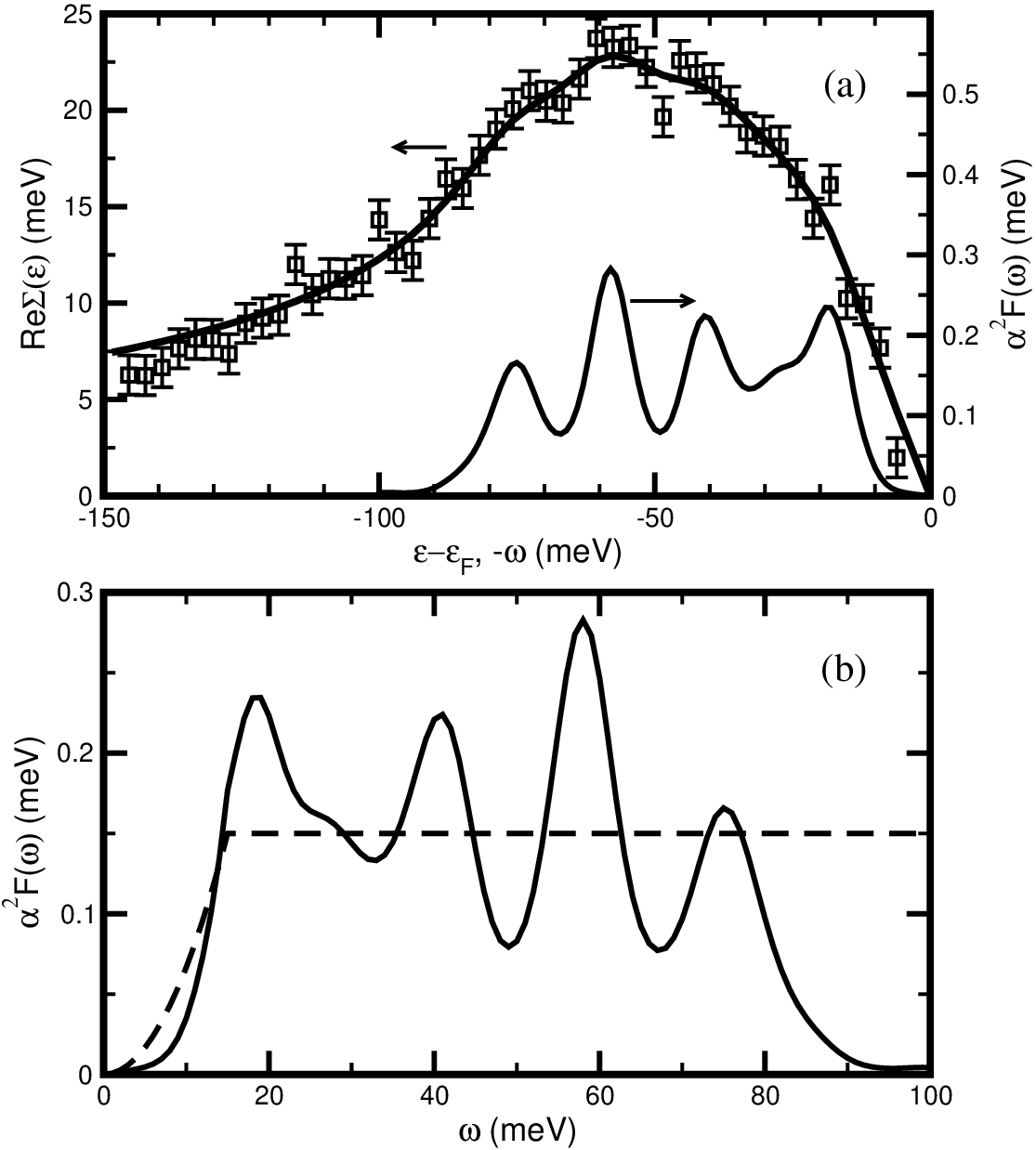}

\caption{\label{cap:Result}(a) Real part of the self energy. Solid
line shows the MEM fitting. The error bars
$\sigma_{i}=1\,\mathrm{meV}$ are determined from the noise level of
the data. (b) Extracted Eliashberg function. The dashed line shows the
constraint function $m(\omega)$ used: $\omega_{D}=15\,\mathrm{meV}$,
$\omega_{m}=100\,\mathrm{meV}$, and $m_{0}=0.15$ (See
Eq.~\ref{eq:model}). The extracted Eliashberg function is also shown
in (a) to indicate the origin of its structures in the self-energy
(note the x-axis is reversed).}
\end{figure}

Figure~\ref{cap:photoemmission}(a) shows the momentum distribution
curves (MDCs) of the Be(10$\bar{1}$0) S1 surface state. The
quasi-particle dispersion $\epsilon(\mathbf{k})$ is determined from
MDCs by Lorentzian fittings~\cite{Lashell2000}, and is shown in
Fig.~\ref{cap:photoemmission}(b).  The real part of the self energy is
calculated by
$\mathrm{Re}\Sigma(\epsilon,\hat{\mathbf{k}})=\epsilon(\mathbf{k})-\epsilon_{0}(\mathbf{k})$,
where $\epsilon_{0}(\mathbf{k})$ is the bare quasi-particle
dispersion, i.e., without EPC. Within the small energy scale
considered, it is sufficient to approximate
$\epsilon_{0}(\mathbf{k})\approx-\hbar\upsilon_{F}(k-k_{F})+\beta(k-k_{F})^{2}$.
Following the extrapolation procedure that will be detailed later, we
find that $\hbar\upsilon_{F}=2.95\,\mathrm{eV}\cdot\mathrm{\AA}$ and
$\beta=6.74\,\mathrm{eV}\cdot\mathrm{\AA}^{2}$ provide the best fit to
the data, as shown in Fig.~\ref{cap:photoemmission}(b). The resulting
real part of the self energy for the S1 state is shown in
Fig.~\ref{cap:Result}(a).

The real part of the self energy is related to the Eliashberg function
by~\cite{Grimvall}: 
\begin{equation}
\mathrm{Re}\Sigma(\epsilon,\mathbf{\hat{k}};T)=\int_{0}^{\infty}d\omega\alpha^{2}F(\omega;\epsilon,\mathbf{\hat{k}})K\left(\frac{\epsilon}{kT},\frac{\omega}{kT}\right)\,,\label{eq:ReSigma}
\end{equation}
where
$K(y,y^{\prime})=\int_{-\infty}^{\infty}dxf(x-y)2y^{\prime}/(x^{2}-y^{\prime2})$
with $f(x)$ being the Fermi distribution
function. Eq.~\ref{eq:ReSigma} is valid for a normal Fermi liquid
coupled with boson field(s). For a normal metal that has no sharp
electronic structure around the Fermi surface within the small energy
scale of the order of the phonon bandwidth (80~meV), dependence on the
initial state energy $\epsilon$ can be ignored because
$d\alpha^{2}F(\omega;\epsilon,\hat{\mathbf{k}})/d\omega\gg
d\alpha^{2}F(\omega;\epsilon,\hat{\mathbf{k}})/d\epsilon$~\cite{Grimvall}.
Thus, $\alpha^{2}F(\omega;\epsilon,\hat{\mathbf{k}})$ in
Eq.~\ref{eq:ReSigma} can be replaced by the Eliashberg function at the
Fermi level $\alpha^{2}F(\omega)$, defined as
$\alpha^{2}F(\omega)\equiv\alpha^{2}F(\omega;\epsilon_{F},\hat{\mathbf{k}})$.

With this simplification, Eq.~\ref{eq:ReSigma} poses an integral
inversion problem for extracting $\alpha^{2}F(\omega)$ from the real
part of the self energy. The most straightforward way to do the
inversion is the least-squares method that minimizes the
functional:
\begin{equation}
\chi^{2}=\sum_{i=1}^{N_{D}}\frac{\left[D_{i}-\mathrm{Re}\Sigma(\epsilon_{i})\right]^{2}}{\sigma_{i}^{2}}\,,\label{eq:chi}
\end{equation}
against the Eliashberg function $\alpha^{2}F(\omega)$, where $D_{i}$
are the data for the real part of the self energy at energy
$\epsilon_{i}$; $\mathrm{Re}\Sigma(\epsilon_{i})$ is defined by
Eq.~\ref{eq:ReSigma} and is a functional of $\alpha^{2}F(\omega)$;
$\sigma_{i}$ are the error bars of the data; and $N_{D}$ is the total
number of data points.  Unfortunately, such a straightforward approach
fails because the inversion problem defined by Eq.~\ref{eq:ReSigma} is
unstable mathematically and the direct inversion tends to
exponentially amplify the high-frequency data noise, resulting in
unphysical fluctuations and negative values in the extracted
Eliashberg function.

To overcome the numerical instability in the direct inversion, one
needs to incorporate the physical constraints into the fitting
process.  The most important constraint is that the Eliashberg
function must be positive. To do this, we employ the Maximum Entropy
Method (MEM)~\cite{Gubernatis1991}, which minimizes the
functional:
\begin{equation}
L=\frac{\chi^{2}}{2}-aS\,\label{eq:L}
\end{equation} 
where $\chi^{2}$ is defined in Eq.~\ref{eq:chi}, and $S$ is the
generalized Shannon-Jaynes entropy,
\begin{equation}
S=\int_{0}^{\infty}d\omega\left[\alpha^{2}F(\omega)-m(\omega)-\alpha^{2}F(\omega)\ln\frac{\alpha^{2}F(\omega)}{m(\omega)}\right]\,.\label{eq:S}
\end{equation}
The entropy term imposes physical constraints to the fitting and is
maximized when $\alpha^{2}F(\omega)=m(\omega)$, where $m(\omega)$ is
the constraint function and should reflect our best \emph{a-priori}
knowledge for the Eliashberg function of the specific system. In this
study, we use the following generic form: 
\begin{equation}
m(\omega)=\left\{ \begin{array}{ll}
m_{0}\left(\omega/\omega_{D}\right)^{2}\,, & \omega\le\omega_{D}\\
m_{0}\,, & \omega_{D}<\omega\le\omega_{m}\\ 0\,, &
\omega>\omega_{m}\end{array}\right.\,,\label{eq:model}
\end{equation}
which encodes the basic physical constraints of the Eliashberg
function: (a) it is positive; (b) it vanishes at the limit
$\omega\rightarrow0$; (c) it vanishes above the maximal phonon
frequency.

The multiplier $a$ in Eq.~\ref{eq:L} is a determinative parameter that
controls how close the fitting should follow the data while not
violating the physical constraints. When $a$ is small, the fitting
will follow the data as closely as possible, and when $a$ is large,
the extracted Eliashberg function will not deviate much from
$m(\omega)$.  There exist a number of schemes (eg. historic, classic,
Bryan's method, etc.) that choose the optimal value of $a$ based on
the data and the constraint function
$m(\omega)$~\cite{Gubernatis1991}. In this study, the classic method
is used.

To minimize Eq.~\ref{eq:L}, Eqs.~\ref{eq:ReSigma} and \ref{eq:S} are
discretized using the iterated trapezoid rule for the integral over
$\omega$, and $\alpha^{2}F(\omega)$ is optimized using a Newton
algorithm developed by Skilling and collaborators that searches along
three prescribed directions in each iterative
step~\cite{Gubernatis1991}.  Detailed reviews of the algorithms can be
found in Ref.~\onlinecite{Gubernatis1991}.

Figure~\ref{cap:Result}(b) shows the extracted Eliashberg function of
the Be(10$\bar{1}$0) S1 surface state and the constraint function
$m(\omega)$. It resolves a number of peaks approximately at 40, 60,
and 75~meV. Low energy modes with $\omega<30\,\mathrm{meV}$ are also
evident. When compared with the first-principles calculations of the
phonon dispersion of this system~\cite{Lazzeri2000}, the resolved
peaks correspond well to those surface phonon modes. The extracted
Eliashberg function automatically cuts off at $\sim80\,\mathrm{meV}$,
which is also consistent with the calculation~\cite{Lazzeri2000}.  The
mass enhancement factor $\lambda$ is calculated by
Eq.~\ref{eq:integral lambda}, yielding a value $0.68\pm0.08$,
consistent with $\lambda=0.65$ obtained in previous measurements using
temperature-dependent line shapes~\cite{Tang2002}.

\begin{figure}
\includegraphics[%
  width=1.0\columnwidth]{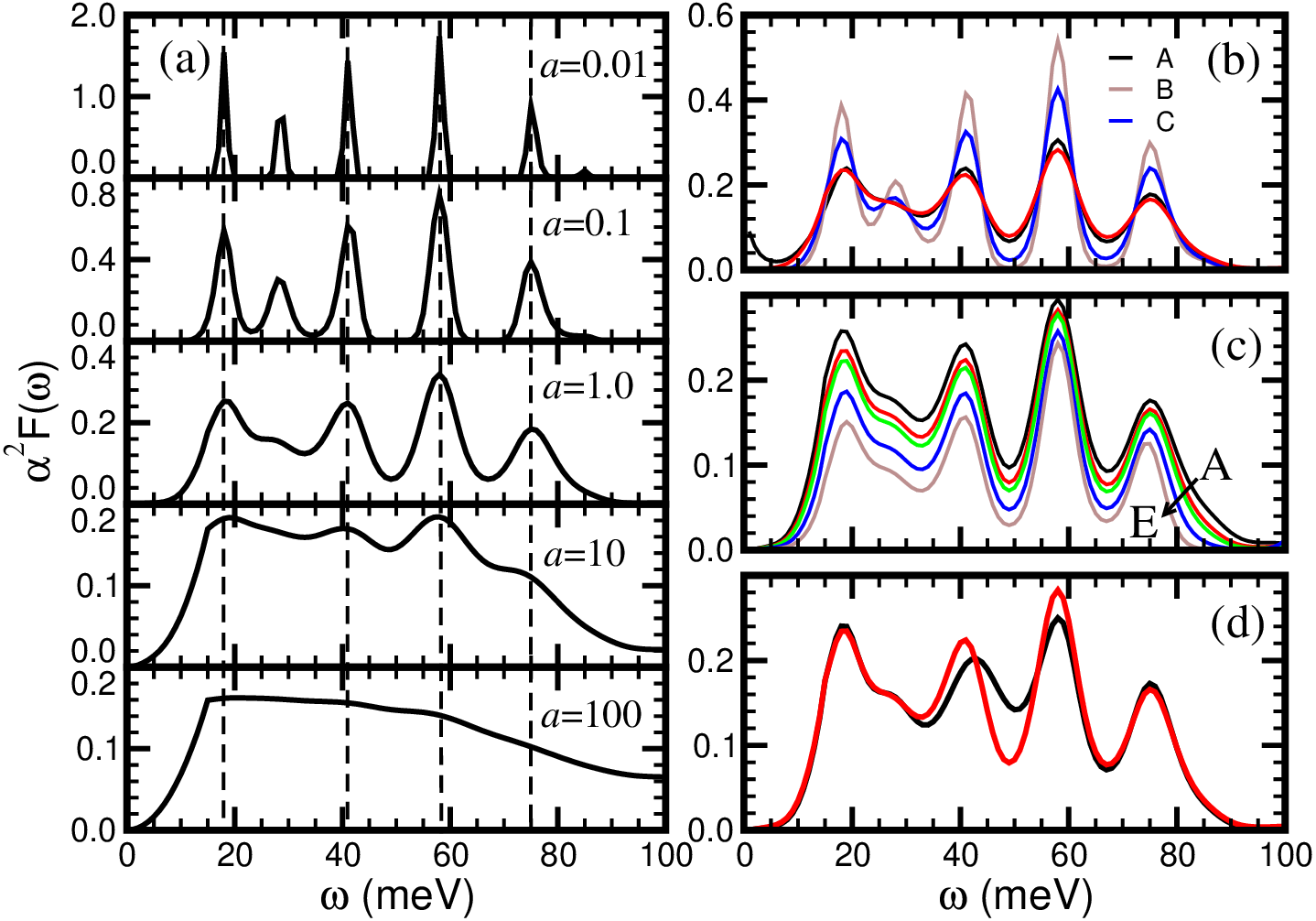}

\caption{\label{cap:Robustness} Various robustness tests to the
fittings.  The red line in each figure indicates the nominal result as
presented in Fig.~\ref{cap:Result}(b). (a) Extracted Eliashberg
functions for different values of $a$. (b) Extracted Eliashberg
functions for different constraint function parameters. For each
curve, one of parameters used in Fig.~\ref{cap:Result}(b) is changed:
A. $\omega_{D}\Rightarrow0$;
B. $\omega_{m}\Rightarrow200\,\mathrm{meV}$; C. $m_{0}\Rightarrow0.3$.
The corresponding values of $a$ determined by the classic method are:
A. $a=1.13;$ B. $a=0.22$; C. $a=0.43$. The nominal result has
$a=1.44$. (c) Extracted Eliashberg functions of the fittings A--E
shown in Fig.~\ref{cap:photoemmission}(b) for different bare particle
dispersions. (d) Extracted Eliashberg function (black line) after
dropping the abnormal data point at $\sim47\,\mathrm{meV}$ in
Fig.~\ref{cap:Result}(a). }
\end{figure}

We have carried out systematic tests to assess the effects of various
parameters involved in the fitting and the robustness of the procedure
against data imperfections. These tests are summarized in
Fig.~\ref{cap:Robustness}.  Based on these tests, which will be
detailed below, we conclude that the fitting procedure can be well
controlled to provide reliable physical insights.

The most important fitting parameter is the multiplier
$a$. Figure~\ref{cap:Robustness}(a) shows the fitting results as a
function of $a$. Changing the value of $a$ does not change qualitative
features of the extracted Eliashberg function such as the number and
the positions of the peaks, although the quantitative changes of the
{}``contrast'' are evident. Furthermore, the estimate of $\lambda$ is
not sensitive to the value of $a$: for $a$ changing from $0.01$ to
$100$, $\lambda$ varies only between $0.64$ and $0.69$.

The classic method determines the optimal value of $a$ based on the
data and the constraint function. Inevitably, the parameters $m_{0}$,
$\omega_{R}$, and $\omega_{m}$ in the constraint function influence
the decision of classic method, as demonstrated in
Fig.~\ref{cap:Robustness}(b).  A proper choice of these parameters is
important to ensure that the algorithm makes the correct decision. In
Fig.~\ref{cap:Result}(b), $m_{0}$ is roughly the average height of the
Eliashberg function, $\omega_{m}$ is slightly higher than the maximal
phonon frequency, and a small but nonzero value of $\omega_{D}$ is
chosen to suppress the artifact near the zero frequency as seen in in
Fig.~\ref{cap:Robustness}(b) for $\omega_{D}=0$. In this way, we have
a constraint function that is close enough to the real Eliashberg
function and is still sufficiently structureless for an unbiased
fitting.

The parameters in the bare particle dispersion, $\upsilon_{F}$ and
$\beta$, are determined from extrapolation to the higher initial state
energy. $\upsilon_{F}$ and $\beta$ are inter-dependent because given a
value of $\beta$, there is only one value of $\upsilon_{F}$ (within a
small window) that can yield the real part of the self energy which
has the correct asymptotic behavior and can be fitted by
Eq.~\ref{eq:ReSigma}.  To find the optimal bare quasi-particle
dispersion, we tried a number pairs of ($\upsilon_{F}$, $\beta$) to
generate the real part of the self energy, then ran the MEM fitting
within $\epsilon<150\,\mathrm{meV}$.  The optimal pair of
($\upsilon_{F}$, $\beta$) is the one that provides the best fit to the
dispersion data in the larger energy window (300~meV).  This procedure
is demonstrated in Fig.~\ref{cap:photoemmission}(b).  The
corresponding extracted Eliashberg functions for different values of
($\upsilon_{F}$, $\beta$) are shown in Fig.~\ref{cap:Robustness}(c).
It can be seen that fittings with less optimal values of
($\upsilon_{F}$, $\beta$) yield result rather close to the nominal
one. This removes the need for high accuracy in determining the bare
quasi-particle dispersion.

The MEM fitting is rather robust against the data imperfections such
as accidental data anomaly, as demonstrated in
Fig.~\ref{cap:Robustness}(d).  The robustness has two origins: (a)
physical constraints built in the fitting automatically filter out
those unphysical data fluctuations; (b) although one data point may be
anomalous, it is statistically less likely that a number of data
points become anomalous simultaneously in a coherent way.

\begin{figure}
\includegraphics[%
  clip,
  width=0.90\columnwidth,
  keepaspectratio]{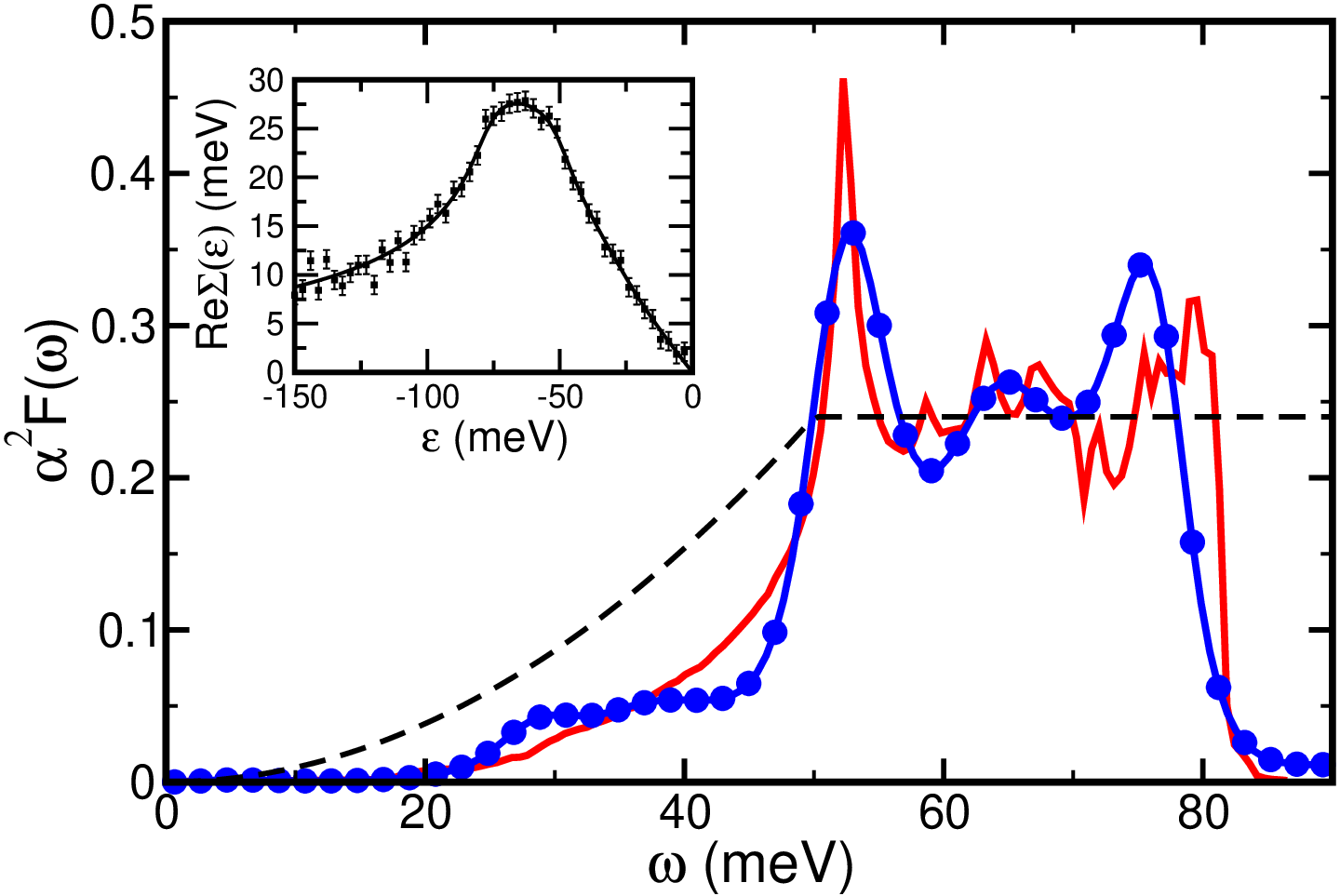}

\caption{\label{cap:testcase} (Color online) A test to the data
extraction procedure. Solid line shows the pre-defined Eliashberg
function. The filled circle-line shows the extracted Eliashberg
function. Dashed line shows $m(\omega)$ with parameters
$\omega_{D}=50\,\mathrm{meV}$, $\omega_{m}=100\,\mathrm{meV}$, and
$m_{0}=0.2$4. Inset shows the data presented to the program and the
MEM fitting. Gaussian noise with $\sigma_{i}=1\,\mathrm{meV}$ has been
added to the self-energy data. $T=30\,\mathrm{K}$. }
\end{figure}

Figure~\ref{cap:testcase} provides a further test on the robustness of
the MEM fitting. Here, we generate a set of data of the real part of
the self energy from a pre-defined Eliashberg function. Gaussian noise
is added to simulate the realistic experimental situation. Figure
\ref{cap:testcase} shows a typical result of the test. It can be seen
that the MEM successfully extracts the overall qualitative features of
the predefined Eliashberg function although the data are rather
noisy. The mass enhancement factor calculated from
Eq.~\ref{eq:integral lambda} is $0.35\pm0.05$, which is very close to
the exact value $0.35$.

Our MEM fitting scheme can be further improved by using a better
constraint function $m(\omega)$. For instance, if the phonon density
of states $F(\omega)$ is known from other reliable sources, one can
use $m(\omega)=\alpha_{0}^{2}\times F(\omega)$ with $\alpha_{0}^{2}$
being determined by optimizing its posterior
probability~\cite{Gubernatis1991}. Such a scheme should allow a
seamless integration of the existing knowledge of the fine spectral
structure and the newly extracted coupling strengths.

Our systematic procedure to extract the Eliashberg function has a
number of advantages. (a) ARPES has much wider applicability than the
traditional method using the single-particle tunneling
characteristics~\cite{Grimvall}.  (b) ARPES allows measurements along
different directions, so the Eliashberg function
$\alpha^{2}F(\omega;\epsilon,\hat{\mathbf{k}})$ on the whole Fermi
surface could be determined. (c) Equation~\ref{eq:ReSigma}, the
theoretical basis of our method, makes only minimal assumption on the
nature of the system, i.e., a normal Fermi liquid without strong
electronic structure near the Fermi energy. (d) The procedure only
utilizes the data for $\mathrm{Re}\Sigma(\epsilon)$, which is easier
to determine and less prone to the imperfections than
$\mathrm{Im}\Sigma(\epsilon)$.

Here, our case study also provides new insights to EPC at the
Be(10$\bar{1}$0) surface. (a) The extracted Eliashberg function is
significantly different from the simple fictitious models (e.g., Debye
and Einstein models) previously used to interpret the
data~\cite{Lashell2000,Tang2002}; (b) More than 75\% (0.5 out of 0.68)
of the enhanced $\lambda$ is contributed by the low frequency surface
modes below $50\,\mathrm{meV}$ that are not present in the bulk phonon
spectrum. Similar behavior is also observed in Be(0001)
surface~\cite{Eiguren2002,Tang2003}.  This raises the question whether
the low frequency surface phonon modes are responsible for the large
mass enhancement factors observed in many metal
surfaces~\cite{Matzdorf1996,Balasubramanian1998,Hengsberger1999,Valla1999,Lashell2000,Rotenberg2000,Tang2002,Ast2002};
(c) The average phonon frequency~\cite{Kulic2000},
$\ln\omega_{\log}=(2/\lambda)\int_{0}^{\infty}d\omega[\alpha^{2}F(\omega)/\omega]\ln\omega,$
is calculated to be $29\,\mathrm{meV}$, which is substantially smaller
than its bulk value
($\sim60\,\mathrm{meV}$)~\cite{Balasubramanian1998}.  This reduces the
estimated $T_{c}$ for possible superconductivity.

In summary, we have proposed a systematic way to extract the
Eliashberg function from the high-resolution ARPES data. The MEM is
employed to overcome the data imperfections and numerical
instability. By using this new technique, we have provided new
insights to EPC at the Be(10$\bar{1}0)$ surface. We expect the
technique to be useful in many situations, for instance, in the study
of the possible role of EPC in high-temperature superconductivity.

\begin{acknowledgments}
This work was partially supported by NSF DMR--0105232 (EWP),
DMR--0071897 (ZXS), and DMR--0306239 (ZYZ). The experiment was
performed at the ALS of LBNL, which is operated by the DOE's Office of
BES, Division of Materials Sciences and Engineering, with contract
DE-AC03-76SF00098.  The division also provided support for the work at
SSRL with contract DE-FG03-01ER45929-A001. The work was also partially
supported by ONR grant N00014-98-1-0195-P0007, and by Oak Ridge
National Laboratory, managed by UT-Battelle, LLC, for the
U.S. Department of Energy under Contract DE-AC05-00OR22725.
\end{acknowledgments}

\end{document}